\documentclass[twocolumn,showpacs,preprintnumbers,amsmath,amssymb]{revtex4}
\usepackage{graphicx}% Include figure files
\usepackage{dcolumn}% Align table columns on decimal point
\usepackage{bm}% bold math

\begin{document}

\title{Enhanced damping of ion acoustic waves in dense plasmas}% \\

\author{S. Son}
\email{seunghyeonson@gmail.com}
\affiliation{18 Caleb Lane, Princeton, NJ 08540}
\author{Sung Joon Moon\footnote{Current Address: 28 Benjamin Rush Ln. Princeton, NJ 08540}}
\affiliation{Program in Applied and Computational Mathematics, Princeton University, Princeton, NJ 08544}
\date{\today}% It is always \today, today,

\begin{abstract}
A theory for the ion acoustic wave damping in dense plasmas and
warm dense matter, accounting for the Umklapp process, is presented.
A higher decay rate compared to the prediction from the Landau damping theory
is predicted for high-Z dense plasmas where the electron density ranges from
$10^{21}$ to $ 10^{24}~\mathrm{cm^{-3}}$ and 
the electron temperature is moderately higher than the Fermi energy.
\end{abstract}


\maketitle

The ion acoustic wave, a longitudinal collective mode in plasmas, plays
a crucial role in a range of applications, such as the Thomson
scattering~\cite{Glenzer, Chihara}
and the Brillouin scattering~\cite{Murray}. 
Understanding the dynamics of the wave in dense plasmas or warm dense matter
is important in various context, including the inertial confinement
fusion~\cite{Tabak, Drake} and the compression of
x-rays~\cite{fast4, Free2,Fisch, Fisch2}.
%How a plasma wave develops and damps is the most basic dynamical process
%to be understood.

The decay rate of the ion acoustic waves in plasmas is often modeled by
the prevalent Landau damping theory.
However, this theory is inadequate for {\em dense} plasmas, as the Umklapp
process, which is not accounted for, becomes pronounced in high densities.
It was shown that the Umklapp process dominates the Landau damping
for low-$k$ plasmons~\cite{Sturm, sonpre2, Ku}.
Even though the detailed underlying physical mechanisms of the plasmons are
different from those of the ion acoustic waves, it is expected that
%, to some degree,
the Umklapp process is also important to the ion acoustic waves. 
The goal of this paper is to estimate the effect of this process.
Starting from the plasmon damping theory in dense plasmas~\cite{sonpre2}, 
a new theory predicting the ion acoustic wave sampling is proposed and
a regime where the decay rate is larger than the prediction by the Landau
damping theory is identified. 
This result would have implications on the Brillouin scattering of dense
plasmas, the x-ray Thomson scattering,
and the reflection problem in the inertial confinement fusion.

First we provide a brief review on the Landau damping theory for the ion-acoustic wave. 
Only a neutral plasma of single ion-species ions is considered for simplicity. 
We denote the electron (ion) temperature by $T_e$ ($T_i$), the corresponding density by
$n_e$ ($n_i$), the mass by $m_e$ ($m_i$), and the charge by $Z_e = 1$ ($Z_i = Z$),
where the charge neutrality condition reads $n_e = Z n_i$. 
%For a moderate Landau damping of an acoustic wave to occur, the electron temperature
%should exceed the ion temperature, which we assume to be the case below.
The longitudinal dielectric function is
$ \epsilon(\mathbf{k}, \omega) = 1 + \chi_e + \chi_i$,
where 
\begin{equation} 
\chi_{i,e} =  \frac{\omega_{i,e}^2}{k^2}  \int  \frac{ \mathbf{k} \cdot \mathbf{\nabla_v} f_{i,e}}{ \omega - \mathbf{k} \cdot \mathbf{v}} d^3 \mathbf{v}  \mathrm{,}  \label{eq:chi}
\end{equation}
$\omega_{i,e}^2 = 4 \pi n_{i,e} Z_{i,e} e^2 / m_{i,e}$ is the ion (electron) plasmon frequency,
and $f_{i,e} $ is normalized as $\int f_{i,e} = 1 d^{3} \mathbf{v} $.   
We assume  $v_{\mathrm{ti}} < \omega / k \ll v_{\mathrm{te}} $, 
where $v_{\mathrm{ti}, \mathrm{te}} =  \sqrt{T_{\mathrm{i,e}}/ m_{i,e}} $.
This is a necessary condition for a moderate Landau damping. 
Under this assumption,
$\chi_e$ and $\chi_i$ can be estimated to be $\chi_e \cong 1/(k \lambda_{\mathrm{de}})^2 $ and $\chi_i \cong  - (\omega_{i} / \omega)^2$, where $\lambda_{\mathrm{de}} = \sqrt{T_e / 4 \pi n_e e^2}$ is the Debye screening length.
The condition $\epsilon = 0$ yields the dispersion relation for the ion acoustic wave,
\begin{equation}
\omega_{\mathrm{iaw}}^2  = \frac{ k^2 V^2 }{ 1 + (k \lambda_{\mathrm{de}})^2 } \mathrm{,}
\end{equation}
where $ V = \sqrt{ Z T_e / m_i} $.
The ion acoustic wave decay rate from the Landau damping theory is given to be
\begin{equation} 
\gamma / \omega_{\mathrm{iaw}}  = \left( \frac{\omega_{\mathrm{iaw}}}{\omega_i} \right)^2 \mathrm{Im} \left[ \chi_e + \chi_i  \right]\mathrm{,} \label{eq:landau}
\end{equation} 
where 
\begin{eqnarray}
\mathrm{Im} \left[ \chi_{i,e} \right]= \left(\frac{\omega_{i,e}}{k v_{ti, te}}\right)^2  S \frac{1 }{ \sqrt{2 \pi}} \exp(-S^2/2) \mathrm{,}  \nonumber \\ \nonumber 
 \end{eqnarray}
and $ S =  \omega_{\mathrm{iaw}} / k v_{\mathrm{te},\mathrm{ti}} $.
%We consider the situation $ \mathrm{Im} [\chi_i] < \mathrm{Im} [\chi_e ]$,
%which requires the condition $T_e > T_i $.
%From now on, we focus our discussion on $\chi_e$. 

According to the Landau damping theory for the ion acoustic wave,
there are many electrons satisfying the resonance condition, of which
energy is very low compared to the average kinetic energy.
The distribution function around the resonance condition hardly varies,
and nearly even electron population on both sides of the resonance condition
results in little net energy transfer between the wave and the electrons.
In other words, the derivative of the distribution function with respect to
the velocity nearly vanishes at the resonance condition, which makes the
decay rate small. 
This physical picture that electrons are freely-streaming and interacting only
with the wave is no longer accurate in dense plasmas, 
because the distortion in the electron motion due to the presence of the ions
(i.e., the Umklapp process) becomes important~\cite{Sturm, sonpre2}.

The effect of the Umklapp process on the ion acoustic wave damping can be
analyzed by an approach similar to what was taken in developing the plasmon
damping theory for dense plasmas in Ref.~\cite{sonpre2}.
Below we follow nearly the same steps as in Sec. IV therein.
In the presence of the potential of the form
$\phi(\mathbf{x},t) = \phi \exp(i\mathbf{k}\cdot \mathbf{x} -i\omega t) + \phi^* \exp(-i\mathbf{k}\cdot \mathbf{x} +i\omega t )$, the wave packet is modified to be
\begin{eqnarray}
 \parallel\sigma\rangle = |\sigma \rangle  + \sum_{\sigma_1}\frac{ e\phi}{ \hbar \omega - E_{\sigma_1} + E_{\sigma}} |\sigma_1 \rangle  \langle \sigma_1| \exp(i\mathbf{k}\cdot \mathbf{x})| \sigma\rangle  \nonumber \\
+ \sum_{\sigma_1}\frac{ e\phi^*}{ -\hbar \omega + E_{\sigma_1} - E_{\sigma}} 
|\sigma_1 \rangle \langle \sigma_1| \exp(-i\mathbf{k}\cdot \mathbf{x}) |\sigma\rangle  \nonumber \mathrm{,} \\ \label{eq:sigma} \nonumber  
\end{eqnarray}
where the perturbation theory of the first order is used assuming the perturbation is weak,
$|~\rangle$ denotes the original eigenstate, $\parallel~\rangle $ denotes the perturbed eigenstate,
and $\sigma$ is an index for the eigenstate.
Then the perturbation in the density is given to be
\begin{eqnarray}
\delta n(k,\omega) = \sum_{\sigma_1} f(\sigma_1) |\langle\sigma_1 \parallel \exp(i\mathbf{k}\cdot\mathbf{x}) \parallel\sigma_1 \rangle|^2 \nonumber \\ 
= \sum_{\sigma_1, \sigma_2} e\phi  (f(\sigma_1) -f(\sigma_2)) \beta(\sigma_1, \sigma_2, \mathbf{k}, \omega ) \mathrm{,} \nonumber \\  \nonumber 
\end{eqnarray}
where $f(\sigma_i)$ is the occupation number, and $\beta(\sigma_1, \sigma_2)$ is
\begin{eqnarray}
  \beta(\sigma_1, \sigma_2) =
\frac{\langle\sigma_1|\exp(-i\mathbf{k}\cdot \mathbf{x})|\sigma_2\rangle\langle\sigma_2|\exp(i\mathbf{k}\cdot \mathbf{x})|\sigma_1\rangle}{ \hbar \omega - E_{\sigma_1} + E_{\sigma_2}}  \mathrm{.} \nonumber  \\ \nonumber 
\end{eqnarray}
$\chi_e$ is, up to the first order in $\phi$,
\begin{eqnarray}
 \chi_e = \frac{4\pi n_e e^2}{k^2} \sum_{\sigma_1, \sigma_2} e\phi  (f(\sigma_1) -f(\sigma_2)) \nonumber \\ \times  
\frac{\langle\sigma_1|\exp(-i\mathbf{k}\cdot \mathbf{x})|\sigma_2\rangle\langle\sigma_2|\exp(i\mathbf{k}\cdot \mathbf{x})|\sigma_1\rangle}{ \hbar \omega - E_{\sigma_1} + E_{\sigma_2}} \mathrm{.} \label{eq:adler}
\end{eqnarray}
With an appropriate choice of the eigenstates under a given condition, Eq.~(\ref{eq:adler})
can be applied to various situations.
For example, Sturm~\cite{Sturm} used the eigenstate
$ |\sigma\rangle = |\mathbf{q}\rangle + \sum_{\mathbf{q}_1 \neq \mathbf{q}} |\mathbf{q}_1 \rangle \langle \mathbf{q}_1 | V |\mathbf{q}\rangle/ ( E_{\mathbf{q}} - E_{\mathbf{q}_1})$, 
where $\mathbf{q}$ and $\mathbf{q}_1$ are the wave vectors,
and $\langle \mathbf{q}_1 | V |\mathbf{q}\rangle$ is the pseudo-potential.
We will suggest an appropriate eigenstate for the ion acoustic wave,
after discussing the ion dynamics below.
%This formalism has been applied to  dense plasmas or warm dense matter to obtain the effect of the ions on the plasmon decay.  

\begin{figure}
\scalebox{0.3}{
\includegraphics{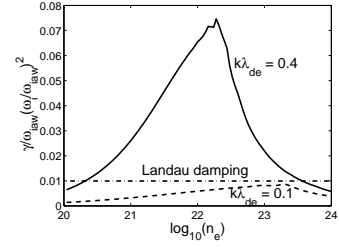}}
\caption{\label{fig:1} 
The decay rates computed by the theory developed in this paper,
compared with the prediction of the Landau damping theory (dot-dashed line)
which is almost constant for a range of the wave vector.
The parameters used are $m_i =26$, $Z = 3$, $T_i = 20 \ \mathrm{eV}$, and $T_e = 100 \ \mathrm{eV}$.
Two cases of the ion acoustic wave vectors are shown, $k=0.1 \ k_{\mathrm{de}}$ and $k=0.4 \ k_{\mathrm{de}}$.
%The y-axis is  $\gamma / \omega_{\mathrm{iaw}} (\omega_i/\omega_{\mathrm{iaw}})^2$ and the x-axis is $\log_{10}(n_e) $,
$n_e $ is in the unit of $\mathrm{cm}^{-3}$. 
%The x-axis is $log_{10}(n_e)$ and the y-axis is $\gamma / \omega_{mathrm{iaw}}$. 
}
\end{figure}

Assuming each ion is located at $\mathbf{X}_i$ in the time scale
of the electron damping on the wave (i.e., the Born-Oppenheimer approximation),
we compute the damping rate of the ion-acoustic wave due to the electrons in the presence
of the spatially fixed ions,
%we compute the dynamical property of the electrons in the presence of the fixed ions,
and then obtain the average dynamics by integrating over the probability distribution
of $\mathbf{X}_i$.
Without loss of generality, $\mathbf{X}_i$ can be assumed to follow
the correlation average,
\begin{equation}
\langle \sum_{i,j}\frac{\exp(i\mathbf{s}\cdot (\mathbf{X}_i-\mathbf{X}_j))}{V_c} \rangle= n_I(\mathbf{s})  \mathrm{,} \label{eq:I}
\end{equation}
where $V_c$ is the volume of the region under consideration, and $n_I(\mathbf{s})$ is the static
two-point correlation function of the ions.
%For independent ions, $n_I(\mathbf{s}) = n_I$. 
In the presence of spatially fixed ions, the electron's free wave eigenfunction is modified to be 
\begin{equation}
|\sigma\rangle  =  |\mathbf{q}\rangle + \sum_{i} \int \frac{d^3 \mathbf{q}_1}{(2\pi)^3} \frac{\exp(-i\mathbf{\mathbf{q}}_1\cdot \mathbf{X}_i) U(\mathbf{q}-\mathbf{q}_1)}{ E_{eff}(\mathbf{q}, \mathbf{q}_1)} |\mathbf{q}_1\rangle \mathrm{,}
\label{eq:eigen2}
\end{equation} 
where $U(\mathbf{q}) = 4 \pi Z e^2/ (|\mathbf{q}|^2 + k^2_{\mathrm{de}}) $ is
the Fourier transform of the screened ion-electron potential.  

For the plasmon damping, $ E_{eff}(\mathbf{q}, \mathbf{q}_1) =  E(\mathbf{q}) - E(\mathbf{q}_1) $
can be used, as was used by Sturm~\cite{Sturm}.  
This is a good approximation for the plasmons, as the plasmon energy $\hbar \omega_{e}$
is very high and the non-degenerate perturbation theory is good enough for the calculation
of the distortion of the electron wave packet due to the presence of ions. 
However, in the case of the ion acoustic waves, the wave energy
$\hbar \omega_{\mathrm{iaw}}$ is much smaller than the electron temperature, 
and the perturbation theory is almost degenerate.
In order to account for this effect,
we choose $E_{eff}(\mathbf{q}, \mathbf{q}_1) =  E(\mathbf{q})  + | E(\mathbf{q}) - E(\mathbf{q}_1)|$,
which is a good approximation for the nearly elastic or large inelastic scattering.
The susceptibility obtained from Eqs.~(\ref{eq:adler}), (\ref{eq:I}) and  (\ref{eq:eigen2}) is
\begin{equation}
 \chi_e(\mathbf{k},\omega) = \chi_{\mathrm{rpa}}(\mathbf{k},\omega) + \chi_{\mathrm{dense}}(\mathbf{k},\omega) \mathrm{,} \nonumber 
\end{equation} 
where $\chi_{\mathrm{rpa}}$ is given in Eq.~(\ref{eq:chi}),
and the subscript stands for the random phase approximation. 
% quantum-mechanically given as 
%\begin{equation}
%\chi_{\mathrm{rpa}} = (\frac{4\pi e^2}{k^2})\int \frac{d^3 \mathbf{q}}{ (2\pi)^3 } \frac{f(\mathbf{k}+\mathbf{q}) - f(\mathbf{q}) }{ \hbar \omega - E(\mathbf{k}+\mathbf{q}) + E(\mathbf{k}) } \mathrm{,} \label{eq:lind}
%\end{equation}
%and 
$\chi_{\mathrm{dense}}$ is of our main interest, which is given to be 
\begin{eqnarray}
\chi_{\mathrm{dense}}(\mathbf{k}, \omega) =\frac{4\pi Z e^2}{k^2} \int \frac{d^3 \mathbf{s}}{(2\pi)^3} n_I(\mathbf{s})  \nonumber \\ \nonumber \\ 
\times  \int \frac{d^3 \mathbf{q}}{ (2\pi)^3} \frac{U^2(\mathbf{s})}{A^2(\mathbf{q},\mathbf{k},\mathbf{s})}  
 \frac{ f(\mathbf{q}+\mathbf{k}+\mathbf{s}) - f(\mathbf{q})}{ \hbar \omega  - E(\mathbf{q}+\mathbf{k}+\mathbf{s}) + E(\mathbf{q})} \mathrm{,} \nonumber \\
\label{eq:real}
\end{eqnarray}
where $A(\mathbf{q},\mathbf{k},\mathbf{s})$ is
\begin{eqnarray}
A^{-1}(\mathbf{q},\mathbf{k},\mathbf{s}) = \frac{1}{E_{eff} (\mathbf{q}, \mathbf{q}+\mathbf{s})}
 - \frac{1}{ E_{eff}(\mathbf{q}+\mathbf{k}, \mathbf{q}+\mathbf{k}+\mathbf{s})} \mathrm{.}  \nonumber  \\ \nonumber 
 \end{eqnarray}
The imaginary part of $\chi_{\mathrm{dense}}$ can be obtained from Eq.~(\ref{eq:real})
by replacing   the  denominator, $1/(\hbar \omega  - E(\mathbf{q}+\mathbf{k}+\mathbf{s}) + E(\mathbf{q}))$,  by the delta function  $ \pi \delta ( \hbar \omega  - E(\mathbf{q}+\mathbf{k}+\mathbf{s}) + E(\mathbf{q})) $.  
%by replacing the delta function  $ \pi \delta ( \hbar \omega  - E(\mathbf{q}+\mathbf{k}+\mathbf{s}) + E(\mathbf{q})) $ in the  denominator by $\hbar \omega  - E(\mathbf{q}+\mathbf{k}+\mathbf{s}) + E(\mathbf{q})$.
%\begin{eqnarray}
%\mathrm{Im}\left[ \chi_{\mathrm{dense}}\right] =  (\frac{4\pi e^2}{k^2})
%\int \frac{d^3 \mathbf{s}}{(2\pi)^3} n_I(\mathbf{s})  \int \frac{d^3 \mathbf{q}}{ (2\pi)^3}\frac{U^2(\mathbf{s})}{A^2(\mathbf{q},\mathbf{k},\mathbf{s})}  \nonumber \\ \nonumber \\
%\times  \pi \delta ( \hbar \omega  - E(\mathbf{q}+\mathbf{k}+\mathbf{s}) + E(\mathbf{q})) \nonumber \\ \nonumber  
%\times \left(f(\mathbf{q}+\mathbf{k}+\mathbf{s}) - f(\mathbf{q})\right)\mathrm{.} \nonumber \\ \nonumber \\
% \label{eq:eigen4}
%\end{eqnarray}
For a Maxwellian plasma, $\mathrm{Im} [\chi_{\mathrm{dense}}] $ can be
further simplified using the velocity integration $ \mathbf{v} = \hbar \mathbf{q} / m_e$,
instead of the wave vector $\mathbf{q}$: 
\begin{eqnarray}
\mathrm{Im}\left[ \chi_{\mathrm{dense}}\right] = \int \frac{n_I(\mathbf{s}) d^3 \mathbf{s}}{ (2\pi)^3}   \left(\frac{4\pi Z e^2}{k^2 + k_{\mathrm{de}}^2}\right)^2 \left(\frac{\omega_{e}^2}{k^2}\right) \nonumber \\ \nonumber 
\times \int d^3 \mathbf{v} \left(A^{-2}( \mathbf{v}, \mathbf{k}, \mathbf{s}) \pi \delta(\omega - (\mathbf{k} + \mathbf{s})\cdot \mathbf{v}) \right)   \nonumber \\ \nonumber \\ 
\times  \frac{f_M(\mathbf{v} + \frac{\hbar (\mathbf{k} + \mathbf{s}) }{2m_e}) - f_M(\mathbf{v} - \frac{\hbar (\mathbf{k} + \mathbf{s}) }{2m_e})}{\frac{\hbar}{m_e}}  \mathrm{,}   
\label{eq:final}
\end{eqnarray}
where $f_M$ is the Maxwellian distribution of the electron temperature $T_e$,
satisfying the normalization condition $\int f_M d^3 \mathbf{v} = 1 $.
$A^{-1}( \mathbf{v}, \mathbf{k}, \mathbf{s})$ is given to be 
\begin{eqnarray}
A^{-1}(\mathbf{v},\mathbf{k},\mathbf{s}) = 
\frac{1}{E(\mathbf{v}-\frac{\mathbf{k}+\mathbf{s}}{2}) - |
E(\mathbf{v}-\frac{\mathbf{k}+\mathbf{s}}{2}) - E(\mathbf{v}-\frac{\mathbf{k}-\mathbf{s}}{2}) |}\nonumber \\
 -\frac{1}{E(\mathbf{v}+\frac{\mathbf{k}-\mathbf{s}}{2}) - |
E(\mathbf{v}+\frac{\mathbf{k}-\mathbf{s}}{2}) - E(\mathbf{v}+\frac{\mathbf{k}+\mathbf{s}}{2}) |} \nonumber \\ \nonumber 
 \end{eqnarray}
where $\mathbf{s} $ can be integrated in the spherical coordinate system.
For a given set of $\mathbf{k}$ and $\mathbf{s}$, the integration over the velocity
can be reduced to a two-dimensional integral, as one variable is eliminated by the delta function.
%The velocity can be represented as $\mathbf{v} = v_1 \hat{\mathbf{q}}_1 + v_2 \hat{\mathbf{q}}_2 + v_3 \hat{\mathbf{q}}_3 $, where $ \hat{\mathbf{q}}_1 = (\mathbf{k} + \mathbf{s} ) / |\mathbf{k}+\mathbf{s}|$, $\hat{\mathbf{q}}_3 = \mathbf{k} \times \mathbf{s} / |\mathbf{k} \times \mathbf{s}|$, and $\hat{\mathbf{q}}_2 = \hat{\mathbf{q}}_1 \times \hat{\mathbf{q}}_3$.
%$v_1$ becomes fixed by the delta function as $v_1 = \omega_{\mathrm{iaw}}/ |\mathbf{k}+\mathbf{s}|$,
%and the $v_2$ and $v_3$ integrations need to be done numerically. 
The decay rates given by Eqs.~(\ref{eq:landau}) and (\ref{eq:final})
(Fig.~\ref{fig:1}) exhibit that the newly computed decay rate is much higher than
what is given by the Landau damping theory when $k\lambda_{\mathrm{de}} = 0.4$. 
Similar integrations for various physical parameters show that the regime
where the Umklapp process is important has the electron density ranging
from $10^{21}$ to $10^{24}\ \mathrm{cm^{-3}}$.
The high ion charge would even further enhance the decay rate.
In the presence of high-Z ions, the electrons interact more strongly
with the ion acoustic wave, the ions being used as the momentum storage. 

%   which 
%for the hydrogen plasma and the plasma with $Z=3 $ and $m_i = 26$
%when $k= 0.1 k_{\mathrm{de}}$ and $k=0.5  k_{\mathrm{de}}$.
%As shown in the figures, the damping rate from the Umklapp process is
%shows that the damping rates attributed to the Umklapp process is
%estimated to be higher than that of the Landau damping theory in some regimes.
%In the presence of ions, the electrons interacts more strongly with the ion
%acoustic wave by exchanging the momentum with the ions.  

In summary, it is suggested that the ion acoustic wave decay rate
could be much higher than the prediction by the Landau damping
theory (see Eq. (\ref{eq:final})).
A rather rough theory, accounting for the effect of the Umklapp process
on the decay, is presented.
%Our computation shows that the decay rates are larger than the prediction
%from the Landau damping theory, especially for high-Z ion plasmas.
%The identified regime, where the Umklapp process is important, has 
%the electron density ranging from $10^{21} \mathrm{cm^{-3}} $ to $10^{24} \mathrm{cm^{-3}}$ and the high ion charge makes the enhanced decay more pronounced because the interaction of electrons with ions  is strong.   
The theory proposed here is far from being complete.
For instance, the perturbation expansion given in Eq. (\ref{eq:eigen2}) may
deviate significantly for the electrons of small kinetic energy (less than 1 eV).
However, it highlights an important weak point of the prevalent
Landau damping theory, when applied to dense plasmas:
%he decay of the ion-acoustic wave mainly arises from the interaction between
%he wave and the freely streaming low-energy electrons.
The ion acoustic wave damping in dense plasmas may arise from the three-party
interaction (electron, ion, and wave), as opposed to the two-party interaction
(electron and wave) the Landau damping theory is based on.

Some comments on the relationship of our theory to the plasma kinetic theory are in order. 
In plasma physics, 
the damping of various waves due to the ion-electron collisions is often considered
in terms of the drag of the electron motion due to ions.
For example, the inverse bremsstrahlung is treated in this way~\cite{Dawson}.
%In principle, the same result can be obtained through a full quantum wave packet
%treatment as shown here; Shima~\cite{Shima} presents the details.
%In the case of the inverse bremsstrahlung, see for details Ref. \cite{Shima}. 
When the classical ion-electron collision picture breaks down,
the dielectric function approach presented here is superior to the classical
kinetic approach as demonstrated in the computation of the plasmon damping
in the solid state physics problems~\cite{Ku,  Pitarke}.
In the regime we consider, the classical electron-ion collisions are no longer valid  
and the quantum mechanical treatment is necessary.
It would be interesting to consider the quantum ion acoustic wave as well~\cite{qiaw},
in the context of the quantum electron degeneracy and diffraction~\cite{sonprl, sonpla, sonlandau}.

%\bibliography{iaw}% Produces the bibliography via BibTeX.

\end{document}